\documentclass[aps,pra,reprint,nofootinbib,amsmath,amssymb,showpacs,floatfix,longbibliography]{revtex4-1}

\usepackage{graphicx,units,multirow}

\newcommand{\ket}[1]{\ensuremath{\vert{#1\rangle}}}

\newcommand{\ketbra}[2]{\ensuremath{|{#1 \rangle}{\langle #2}|}}
\newcommand{\op}[1]{\hat{#1}}
\newcommand{\I}{\text{i}}
\newcommand{\E}{\text{e}}
\providecommand{\abs}[1]{\left\lvert#1\right\rvert}

\usepackage[colorlinks,allcolors=blue,hyperindex,breaklinks]{hyperref}
\begin{document}

\title{Parity-based, bias-free optical quantum random number generation\\ with min-entropy estimation}
\author{Mathew R.\ Coleman}
\author{Kaylin G.\ Ingalls}
\author{John T.\ Kavulich}
\author{Sawyer J.\ Kemmerly}
\author{Nicolas C.\ Salinas}
\author{Efrain Venegas Ramirez}
\author{Maximilian Schlosshauer}
\affiliation{Department of Physics, University of Portland, 5000 North Willamette Boulevard, Portland, Oregon 97203, USA}

\begin{abstract}
We describe the generation of sequences of random bits from the parity of photon counts produced by polarization measurements on a polarization-entangled state. The resulting sequences are bias free, pass the applicable tests in the NIST battery of statistical randomness tests, and are shown to be Borel normal, without the need for experimental calibration stages or postprocessing of the output. Because the photon counts are produced in the course of a measurement of the violation of the Clauser--Horne--Shimony--Holt inequality, we are able to concurrently verify the nonclassical nature of the photon statistics and estimate a lower bound on the min-entropy of the bit-generating source. The rate of bit production in our experiment is around 13~bits/s. \\[.2cm]
Journal reference: \emph{J.\ Opt.\ Soc.\ Am.\ B} {\bf 37}, 2088--2094 (2020),  DOI: \href{https://doi.org/10.1364/JOSAB.392286}{\texttt{10.1364/JOSAB.392286}}
\end{abstract}

\maketitle

\section{Introduction}

Random numbers constitute a valuable resource \cite{Hayes:2001:aa}, with applications ranging from simulations of complex systems  \cite{Motwani:1996:oo,Metropolis:1949:uu} and cryptography \cite{Gisin:2002:ii} to fundamental quantum experiments \cite{Brunner:2014:oo} and information technology \cite{Hayes:2001:aa}. A quantum random number generator (QRNG) \cite{Herrero:2017:kk} produces bits from an indeterministic quantum process. A typical approach consist of performing measurements on quantum systems prepared in a coherent superposition state. This approach makes use of the fact that the outcomes of such measurements are fundamentally unpredictable \cite{Masanes:2006:oo}. Optical systems based on quantum properties of photons are particularly popular due to the ease with which photons can be produced, manipulated, and measured \cite{Herrero:2017:kk}. There are many different types of such optical QRNGs, distinguished by the particular kind of physical principle and measurement exploited in the generation of bits, such as branching paths \cite{Jennewein:2000:zz}, photon-number statistics \cite{Furst:2010:uu,Ren:2011:aa}, time-of-arrival statistics \cite{Dynes:2008:km,Wahl:2011:oo}, vacuum fluctuations \cite{Gabriel:2010:zz,Zhang:2017:zz,Zheng:2019:aa}, and Raman scattering \cite{Collins:2015:uu}.  Systems based on the detection of single photons have shown especially high bit rates and efficiencies \cite{Jennewein:2000:zz,Furst:2010:uu,Ren:2011:aa,Dynes:2008:km,Wahl:2011:oo}. They have also enabled a concurrent verification of quantum randomness and security in terms of bounds on the source min-entropy obtained from quantum state tomography \cite{Fiorentino:2006:lm,Fiorentino:2007:ll}, prepare-and-measure protocols \cite{Lunghi:2015:oo}, and violations of Bell inequalities \cite{Pironio:2010:aa, Pironio:2013:qa,Fehr:2013:kn,Christensen:2013:aa}.  A paradigmatic type of a single-photon QRNG is the branching-path generator \cite{Jennewein:2000:zz}. Such a generator may be realized by letting single photons, each prepared in a superposition state $(\ket{H}+\E^{\I\phi}\ket{V})/\sqrt{2}$, traverse a polarizing beam splitter, followed by their detection at the two outputs of the beam splitter. Each such polarization measurement in the $HV$ basis then produces a single random bit. 

From a fundamental point of view, the fact that a QRNG generates bits from a physically random process makes it superior to a pseudorandom number generator (PRNG). A PRNG uses a deterministic algorithm fed with an initial seed, and thus its output is by definition computable \cite{Calude:2010:kk}. In contrast, for a QRNG the value indefiniteness implied by the Kochen--Specker theorem \cite{Kochen:1967:hu} can guarantee the incomputability of the output \cite{Abbott:2012:kk,Calude:2008:za}. One can also exploit the properties of quantum entanglement to experimentally certify the randomness and privacy (and hence security) of the output of a QRNG \cite{Pironio:2010:aa,Pironio:2013:qa,Fehr:2013:kn}. While these are striking advantages over PRNGs, practical realizations of QRNGs exhibit problems of their own \cite{Herrero:2017:kk}. In particular, they tend to be susceptible to substantial bias and correlation effects that may detrimentally affect the quality of the output from the point of view of statistical and algorithmic randomness \cite{Calude:2010:oo,Martinez:2018:kk,Abbott:2019:uu}. In the photonic scenario just described, bias may arise from sources such as imperfect state preparation with nonequalized amplitudes, a polarizing beam splitter that does not reliably separate $H$ and $V$ polarizations, and imbalanced detectors. Correlations may be caused, for example, by dead-time and afterpulsing effects in the detectors \cite{Kang:2003:aa,Horoshko:2017:hh,Martinez:2018:kk}. In practical implementations, these imperfections often have a significant influence and can be difficult to avoid \cite{Herrero:2017:kk}. 
Mitigation of bias effects generally requires the application of an unbiasing algorithm during postprocessing of the output \cite{Herrero:2017:kk}, and some optical QRNGs also include explicit calibration \cite{Jennewein:2000:zz} and feedback \cite{Martinez:2018:kk} stages to fine-tune and balance detector efficiencies, coupling ratios of beam splitters, and other parameters. 

Here we describe a single-photon experiment in which bias is avoided by using the parity of accumulated photon counts to produce the random bits. In this way, our QRNG does not require calibration or postprocessing to generate random numbers that, as we will show, pass both all applicable statistical randomness tests of the NIST suite \cite{Rukhin:2010:ll} and a test of Borel normality \cite{Calude:1994:hh,Calude:2010:kk}. Borel normality is an important criterion as it taps the realm of algorithmic (rather than statistical) randomness, an area of particular interest for QRNGs. Yet there exist only a few studies to date that have used Borel normality for testing QRNGs and probing the kind of incomputability and algorithmic randomness that in principle sets them apart from PRNGs \cite{Calude:2010:oo,Martinez:2018:kk,Abbott:2019:uu}. 

In the photonic context, the approach of using the parity of photon counts to produce random numbers was previously explored in an experiment by F\"urst \emph{et al.}~\cite{Furst:2010:uu}, in which attenuated LED light was detected with a photomultiplier tube. This setup also became the basis of the commercial generator quRNG \cite{qurng}. Our own experiment uses a rather different, quantum-optical setup, which is based on polarization measurements on entangled photon pairs generated from spontaneous parametric downconversion. A chief motivation for using this kind of setup is that it enables us to concurrently measure a lower bound on the min-entropy of the source and, in this way, to perform an active assessment of the presence of quantum randomness. We do this in two complementary ways, namely, (i) by tomographically measuring the two-photon quantum state and using a theoretical result derived by Fiorentino \emph{et al.}~\cite{Fiorentino:2006:lm,Fiorentino:2007:ll}, and (ii) by measuring the violation of the Clauser--Horne--Shimony--Holt inequality \cite{Clauser:1969:ii} and applying to it a theorem of Pironio \emph{et al.}~\cite{Pironio:2010:aa}.

This integration of parity-based single-photon quantum random generation with a verification of min-entropy and quantum randomness via measurements on entangled states distinguishes our experiment from other optical QRNGs that produce bias-free random numbers without postprocessing, such as those described in Refs.~\cite{Dynes:2008:km,Wei:2009:ii,Furst:2010:uu,Stipcevic:2015:aa,Wang:2015:ll,Nguyen:2018:uu,Zheng:2019:aa}. Our experiment also provides full control over the preparation of the quantum states and the measurement settings, and it does not require any time-tagging of individual photon events. This last property is a consequence of the fact that only the number of registered photons per counting interval is needed to obtain a random bit, and likewise, the randomness verification via measurements of the quantum state and the CHSH inequality requires the observation only of the statistics of photon counts. 

Our paper is organized as follows. In Section~\ref{sec:min-entr-estim} we briefly review the estimation of the source min-entropy from quantum state tomography and from violations of the CHSH inequality. In Section~\ref{sec:experiment} we describe our experimental setup and the generation of the random bits. In Section~\ref{sec:results} we report the measured bounds on the source min-entropy, and describe the results of randomness testing performed on the generated bit sequences. We offer concluding remarks in Section~\ref{sec:discussion}. 

\section{\label{sec:min-entr-estim}Min-entropy estimation}

In our experiment, polarization measurements on the prepared two-photon state serve as a generator of randomness and thus of entropy. We can estimate the entropy produced by this source in terms of the min-entropy \cite{Cachin:1997:uu}. The min-entropy may be thought of as quantifying the effectiveness of any strategy that tries to guess, at first attempt, the most likely output of the source. In the quantum setting, there are two particularly important methods for estimating the source min-entropy. The first is the use of a measurement of the quantum state to quantify the amount of coherence and bias between the two state components that define the randomness-generating measurement outcomes \cite{Fiorentino:2006:lm,Fiorentino:2007:ll,Coleman:2020:aa}. The second method uses the connection between violations of the CHSH inequality and the nonclassicality of the output \cite{Pironio:2010:aa,Pironio:2013:qa,Fehr:2013:kn}. We will use both of these methods, and below shall briefly describe them in turn; for more details, see Refs.~\cite{Fiorentino:2006:lm,Fiorentino:2007:ll,Coleman:2020:aa,Pironio:2010:aa,Pironio:2013:qa,Fehr:2013:kn}.  

For each of the two methods, we obtain a lower bound on the source min-entropy per measurement event. If we were to directly generate a sequence of random bits from those individual events (with each event producing one bit), then according to the meaning of the min-entropy \cite{Chor:1988:ii}, the min-entropy bound would also give the minimum number of uniform (and secure \cite{Fiorentino:2006:lm,Fiorentino:2007:ll,Pironio:2010:aa}) random bits that can be extracted from this sequence. In our case, however, we do not translate the measurement events individually into bits, but instead produce the bits from the parity of the number of many such accumulated events. Therefore, the numerical value of the min-entropy bound cannot be applied to measuring the extractability of randomness from our sequences of bits. Indeed this is not surprising, since the min-entropy  quantifies the randomness of the process, which is that of individual quantum measurements. Nonetheless, what measuring a nonzero min-entropy bound accomplishes as far as our parity-generated sequences are concerned is a confirmation of the presence of quantum randomness in the measured photon counts \cite{Fiorentino:2006:lm,Fiorentino:2007:ll,Coleman:2020:aa,Pironio:2010:aa,Pironio:2013:qa,Fehr:2013:kn}. Since our random numbers are produced from a property of the measured photon events---namely, the parity of their numbers---those bits must also contain elements of quantum randomness. The presence of quantum randomness also relates to the privacy and security of the output: It guarantees that the polarization measurements done by one party (Alice) produce fresh randomness even if an adversary (Eve) has prepared the state Alice is measuring, or if Alice's photons are quantum-correlated with Eve's \cite{Fiorentino:2006:lm,Fiorentino:2007:ll}. 

For the min-entropy bound obtained from knowledge of the quantum state, we make use of a result by Fiorentino \emph{et al.}~\cite{Fiorentino:2006:lm,Fiorentino:2007:ll}, who showed that for photons in a known state $\op{\rho}$, the min-entropy $H_\infty(\op{\rho})$ per $HV$ measurement is bounded from below by
\begin{equation}\label{eq:fio}
H^\text{min}_\infty(\op{\rho}) = -\log_2\left(\frac{1+\sqrt{1-4C^2}}{2} \right),
\end{equation}
where $C$ is the magnitude of the off-diagonal elements of $\op{\rho}$ expressed in the $HV$ basis. In Ref.~\cite{Coleman:2020:aa} we have discussed the connection underlying Eq.~\eqref{eq:fio} between the off-diagonal elements and quantum randomness. To apply Eq.~\eqref{eq:fio} to our experiment, we take the tomographically reconstructed density matrix for the entangled two-photon state, restrict it to the subspace spanned by $\ket{H}\ket{H}$ and $\ket{V}\ket{V}$ representing the $HH$ and $VV$ coincidences that one may use as generators of a random output, and renormalize the matrix elements of the resulting $2\times 2$ subspace density matrix $\op{\rho}_\text{sub}$ such that the diagonal elements add up to 1 \cite{Coleman:2020:aa}.  

For the CHSH-derived min-entropy bound \cite{Pironio:2010:aa,Pironio:2013:qa,Fehr:2013:kn}, we consider the conditional min-entropy, denoted $H_\infty(R\,|\, M)$, which represents the min-entropy of the output $R$ obtained after $N$ measurements, given knowledge of the ``input'' $M$. Here this input $M$ consists of the polarization measurement settings used for the photons in each measurement round. Pironio \emph{et al.}~\cite{Pironio:2010:aa} (see also Refs.~\cite{Pironio:2013:qa,Fehr:2013:kn} for improved proofs) showed that after $N$ measurement rounds, $H_\infty(R\,|\, M)$ is bounded from below by
\begin{multline}\label{eq:pir}
H^\text{min}_\infty(R \,|\, M) = N\left[ 1-\log_2\left( 1 + \sqrt{2-\frac{S^2}{4}}\right)\right] \\ \text{for $S\ge 2$,}
\end{multline}
where $S$ is the CHSH $S$ value \cite{Clauser:1969:ii} estimated from the data collected in the $N$ measurement rounds. Applied to the photonic scenario, the $S$ value is defined as a linear combination of four expectation values,
\begin{align}\label{eq:S}
S &= E(\theta_{A1}, \theta_{B1})+E(\theta_{A1}, \theta_{B2})\notag \\& \quad +E(\theta_{A2}, \theta_{B1})-E(\theta_{A2}, \theta_{B2}),
\end{align}
where $E(\theta_A,\theta_B) = \langle \op{P}_{\theta_A,\theta_B} \rangle$ and $\op{P}_{\theta_A,\theta_B}=\op{P}_A(\theta_A) \otimes \op{P}_B(\theta_B)$ denotes the observable representing a joint linear-polarization measurement on a pair of photons along the directions defined by the angles $\theta_A$ and $\theta_B$. For the maximally entangled state $\ket{\Phi^+} = \frac{1}{\sqrt{2}} \left( \ket{H}\ket{H} + \ket{V}\ket{V}\right)$, the choices
\begin{align}\label{eq:bellsett}
\theta_{A1}&=0^\circ, \quad
\theta_{A2}=45^\circ,  \notag \\ \theta_{B1}&=+22.5^\circ,  \quad \theta_{B2}=-22.5^\circ
\end{align}
lead to the maximum possible value $S=2\sqrt{2}$, while for any local realistic models one has $S \le 2$ \cite{Clauser:1969:ii,Brunner:2014:oo}. In the absence of superluminal signaling,  CHSH-inequality violations show that the output cannot be entirely predetermined \cite{Masanes:2006:oo}. 

A rigorous experimental demonstration of CHSH-certified randomness would require closure of any Bell loopholes \cite{Christensen:2013:aa,Brunner:2014:oo,Larsson:2014:ja,Shalm:2015:oo}, which is beyond the scope of our experiment. In particular, the settings for each measurement would have to be chosen randomly. (In this sense, the CHSH-based randomness-certification protocol of Ref.~\cite{Pironio:2010:aa} may be regarded as a randomness expander, since a small random seed used for the measurement settings is amplified into certifiable, private randomness through the measured CHSH violation \cite{Pironio:2010:aa,Abbott:2012:kk}.) Here our own aim is more modest: We use the measurement of the CHSH violation as an indicator for the presence of quantum randomness and for the nonclassical nature of the measured photon statistics.

\section{\label{sec:experiment}Experiment}

\subsection{Experimental setup}

\begin{figure}
\centering\includegraphics[scale=.9]{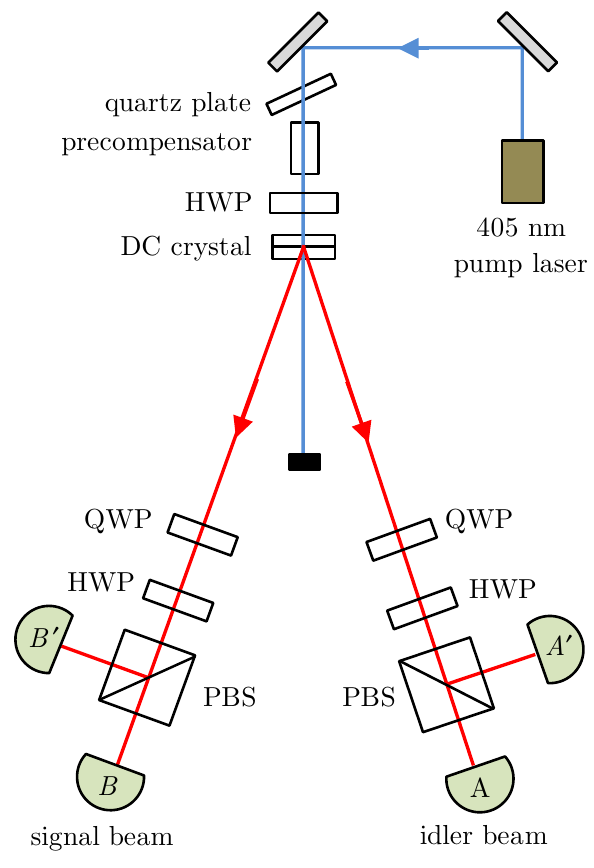}
\caption{\label{fig:setup}Schematic of the experimental setup. DC, downconversion crystal; HWP, half-wave plate; QWP, quarter-wave plate; PBS, polarizing beam splitter. $A$, $A'$, $B$, and $B'$ represent  assemblies containing converging lenses coupled to multimode fiber-optic cables that transmit photons to APD-based single-photon counting modules (not shown).}
\end{figure}

Our experimental setup is schematically shown in Fig.~\ref{fig:setup}. A 405-nm, 150-mW diode laser pumps a pair of closely stacked, 0.5-mm-thick beta-barium borate crystals cut for type-I spontaneous parametric downconversion \cite{Kwiat:1999:oo}. The optic axes of the crystals are oriented at right angles to each other, with one crystal producing pairs of horizontally polarized 810-nm photons, and the other crystal producing vertically polarized pairs. By equally pumping both crystals, we produce a polarization-entangled state resembling $( \ket{H}\ket{H} + \E^{\I \phi} \ket{V}\ket{V}) /\sqrt{2}$. We zero the phase $\phi$ using an X-cut, $\unit[10 \times 10 \times 0.5]{mm}$ quartz plate to generate the Bell state $\ket{\Phi^+} = (\ket{H}\ket{H} + \ket{V}\ket{V}) /\sqrt{2}$. Entanglement quality is enhanced by a $\unit[5 \times 5 \times 5.58]{mm}$ quartz crystal that acts as a precompensator for the walk-off of the orthogonal polarization components inside the downconversion crystal. Signal and idler photons are measured by polarization analyzers composed of quarter-wave plates, half-wave plates, and polarizing beam splitters. Photons emerging from the outputs of the beam splitters are captured by converging lenses coupled to multimode fiber-optic cables and routed to single-photon counting modules based on silicon avalanche photodiodes with a detection efficiency of about 30\% at \unit[810]{nm}. Ambient photons are blocked by 780-nm long-pass filters mounted at the inputs of the photon counting modules. Photons in the signal and idler beams are measured in coincidence within a time window of  about \unit[7]{ns}. Coincidences are processed by a field-programmable gate array \cite{Lord:noyear:uu}. To quantify the closeness of the quantum state generated in our experiment to the Bell state $\op{\rho}_+=\ketbra{\Phi^+}{\Phi^+}$, we tomographically reconstruct the density matrix $\op{\rho}$ for the two-photon state. We calculate the fidelity $F(\rho,\rho_+)=\left(\text{Tr} \sqrt{ \sqrt{\rho}\rho_+\sqrt{\rho}}\right)^2$ and  find $F(\rho, \rho_+) \approx 0.90$. 

\subsection{Bit generation}

Our approach is to carry out a measurement of the violation of the CHSH inequality \cite{Clauser:1969:ii,Brunner:2014:oo} as a verification of the presence of quantum randomness \cite{Pironio:2010:aa, Pironio:2013:qa,Fehr:2013:kn,Christensen:2013:aa}, and to use the same photon counts produced from these polarization measurements to generate the bit sequence. In our experiment, we estimate the $S$ value [see Eq.~\eqref{eq:S}] by measuring the four 2-fold coincidence counts ($AB$, $A'B$, $AB'$, and $A'B'$) at each of the four CHSH measurement settings $(\theta_{A1}, \theta_{B1})$, $(\theta_{A1}, \theta_{B2})$, $(\theta_{A2}, \theta_{B1})$, and $(\theta_{A2}, \theta_{B2})$ given by Eq.~\eqref{eq:bellsett}. For each of these settings, we take $5 \times 10^4$ samples, with each sample accumulated over a counting period of $\tau=\unit[0.2]{s}$. As our experiment does not aim to close loopholes associated with Bell tests \cite{Christensen:2013:aa, Brunner:2014:oo,Larsson:2014:ja,Shalm:2015:oo}, we do not employ random switching of the angles $\theta_A$ and $\theta_B$ between each sample. Instead, we take the full $5 \times 10^4$ samples at a given angle setting and then change one of the angles to proceed to the next setting.

In this way, we obtain a total of $2 \times 10^5$ samples (arising from the $5 \times 10^4$ samples for each of the four angle settings), each containing four 2-fold coincidence counts ($AB$, $A'B$, $AB'$, and $A'B'$). We do not subtract accidental coincidences from the data. We calculate the CHSH $S$ value given by Eq.~\eqref{eq:S} from the measured data and find $S=2.4618 \pm 0.0003$ (the error is estimated from count fluctuations between samples). We produce two different sequences from the measured coincidence count rates. A sequence $\mathbf{x}_1$ containing $2 \times 10^5$ bits is produced from the parity of the number $N(AB)$ of $AB$ coincidence counts in each of the $2 \times 10^5$  samples. A sequence $\mathbf{x}_2$ containing $8 \times 10^5$ bits is produced by using, for each of the $2 \times 10^5$ samples, the parity of each of the four coincidence counts $N(AB)$, $N(A'B)$, $N(AB')$, and $N(A'B')$ to obtain a four-bit string per sample. While this gives a 4-fold increase in the length of the sequence compared to  $\mathbf{x}_1$, one may be concerned that the correlations inherent in the Bell state may introduce detectable correlations into the sequence. We will investigate this question below. Acquisition of the entire data set of $2 \times 10^5$  samples (containing a total of about $3 \times 10^8$ coincidence events) takes about $10^3$ minutes, or about 16.7 hours, and thus sequence $\mathbf{x}_2$ is produced at a rate of \unit[13.3]{bits/s}. [This is somewhat longer than the time one would expect solely based on the length of the counting interval, which is $(2 \times 10^5 \,\text{samples}) \times \unit[0.2]{s} = \unit[4 \times 10^4]{s} \approx \unit[667]{mins}$. The reason for this discrepancy is that after each counting interval of \unit[0.2]{s} is completed, there is a short lag of about \unit[0.1]{s} before the next interval can be counted.] 

\section{\label{sec:results}Results}

\subsection{Min-entropy bounds}

For the min-entropy bound obtained from quantum state tomography using Eq.~\eqref{eq:fio}, in Ref.~\cite{Coleman:2020:aa}, we report the results of the application of this equation to the quantum state prepared in our experiment, and here we shall simply restate these results. We find $C=0.44$ for the magnitude of the off-diagonal elements of the $2 \times 2$ subspace density matrix $\op{\rho}_\text{sub}$ (where $\op{\rho}_\text{sub}$ is obtained from the two-photon density matrix in the manner described in Section~\ref{sec:min-entr-estim}). From Eq.~\eqref{eq:fio}, this gives $H^\text{min}_\infty = 0.44$ per coincidence event. 

For the min-entropy bound obtained obtained from the CHSH violation via Eq.~\eqref{eq:pir}, our measured $S$ value of $S=2.4618 \pm 0.0003$ implies a lower bound on the min-entropy of the output of approximately $0.24$ per joint measurement (coincidence event). The fact that this bound is lower than the bound obtained from tomography may be understood by noting that the scope of these two bounds is quite different. The CHSH-based bound~\eqref{eq:pir} is nonzero only if the CHSH inequality is violated and thus the statistics cannot be modeled classically. In this way, its ultimate aim (in a loophole-free implementation) is the rigorous certification of private, secure randomness in a device-independent fashion \cite{Pironio:2010:aa,Pironio:2013:qa,Fehr:2013:kn}. The tomography-based bound~\eqref{eq:fio}, on the other hand, is nonzero whenever coherence between the two quantum states defining the relevant outcomes is present, and thus it \emph{presumes} the quantum-mechanical tenet that such coherence gives rise to irreducible randomness \cite{Coleman:2020:aa}.

\subsection{\label{sec:randomness-testing}Randomness testing}

\subsubsection{Statistical tests}

\begin{table*}
\centering
\begin{tabular}{|c||ccc|ccc|}
  \hline
\multirow{2}{*}{Statistical Test} & \multicolumn{3}{|c|}{Sequence $\mathbf{x}_1$} & \multicolumn{3}{|c|}{Sequence $\mathbf{x}_2$} \\
&$p$-value & Proportion & $P$-value & $p$-value  & Proportion & $P$-value \\ \hline
Frequency &  0.327383 & 100/100& 0.153763  &  0.90035&99/100  &0.071177\\
Frequency within a Block  &   0.679802 & 100/100 & 0.437274 & 0.263874   &100/100  & 0.045675\\
Runs & 0.880981 & 100/100 & 0.455937 & 0.237737&98/100 &0.595549\\
Longest Run within a Block & 0.230194& 100/100 & 0.153763  & 0.06073 &98/100   &0.759756\\
Cumulative Sums (forward)  & 0.343646&  100/100 & 0.924076 & 0.813301&99/100 &0.137282 \\
Cumulative Sums (backward) & 0.605296&  100/100 & 0.275709  & 0.914125&99/100 &0.062821\\
Discrete Fourier Transform & 0.189098& 100/100 & 0.006196 &0.651679&100/100&0.304126  \\
Serial-1 & 0.510447 & 96/100 & 0.637119 &0.55776 &100/100& 0.224821 \\
Serial-2 & 0.161343 &   98/100 & 0.213309 & 0.318324&99/100&0.383827 \\
Binary Matrix Rank & 0.52999 & n/a & n/a & 0.134194& 20/20 &0.534146 \\ 
Template Matching & 0.894736& 18/20 & 0.534146 & 0.399080& 97/100 & 0.040108\\
Approximate Entropy & 0.070835 & 98/100 & 0.334538 & 0.712837&100/100 &0.595549 \\
Maurer & n/a & n/a &n/a & 0.650933 & n/a & n/a \\
\hline 
\end{tabular}
\caption{\label{fig:nistresults}Results of NIST statistical randomness tests \cite{Rukhin:2010:ll} applied to the sequences $\mathbf{x}_1$ and $\mathbf{x}_2$. Entries marked ``n/a'' indicate that a test could not be applied because of insufficient sequence length.  The Serial test produces two $p$-values as output, shown as ``Serial-1'' and ``Serial-2.'' For $p$-values, the following values for the block length parameter $m$ are used: Frequency within a Block test, $m=2 \times 10^3$ for $\mathbf{x}_1$ and $m=8 \times 10^3$ for $\mathbf{x}_2$; Serial test, $m=15$ for $\mathbf{x}_1$ and $m=16$ for $\mathbf{x}_2$; Approximate Entropy test, $m=10$. For the ``Proportion'' values and $P$-values, the choices for $m$ are: Frequency within a Block test, $m=20$ for $\mathbf{x}_1$ and $m=80$ for $\mathbf{x}_2$; Serial test, $m=8$ for $\mathbf{x}_1$ and $m=10$ for $\mathbf{x}_2$; Approximate Entropy test, $m=5$ for $\mathbf{x}_1$ and $m=7$ for $\mathbf{x}_2$. For the Template Matching applied to $\mathbf{x}_1$ and the Binary Matrix Rank  applied to $\mathbf{x}_2$, a set of only 20 subsequences could be tested due to limited sequence lengths. The Discrete Fourier Transform test has been shown to have problems \cite{Okada:2017:jj} that may render its reliability and sensitivity questionable.}
\end{table*}

We have subjected our generated bit sequences $\mathbf{x}_1$ and $\mathbf{x}_2$ to the statistical randomness tests provided by the NIST suite \cite{Rukhin:2010:ll,Branning:2010:oo}. While such tests cannot positively identify the presence of randomness, they can flag sequences whose statistical patterns do not conform to the patterns expected for a uniform random process. For each NIST test applied to an input sequence, randomness is assessed in terms of a statistical quantity called the $p$-value. This is the probability that an ideal random number generator---such as a generator based on unbiased, independent coin tosses---would have produced a sequence less random (i.e., a sequence that performs worse in the test) than the sequence under inspection, given the type of nonrandom pattern searched for by the test \cite{Rukhin:2010:ll}. For example, if a given test sequence scores a $p$-value of 0.01, then the chance that an ideal random source could have produced a sequence giving a lower (worse) result is 1 in 100. If $p=1$, then the sequence is considered perfectly random with respect to the given test, while $p=0$ would indicate a completely nonrandom sequence. 

Before running the tests, one sets the value of a confidence threshold parameter $\alpha$, and a given sequence is considered to pass the test if its $p$-value is no less than $\alpha$. We use the NIST default value $\alpha=0.01$ unless noted otherwise. First, we apply the tests to the entire sequences $\mathbf{x}_1$ and $\mathbf{x}_2$ to calculate their individual $p$-values. Of the 15 tests provided by the NIST suite, three tests (Linear Complexity, Random Excursions, and Random Excursions Variant) cannot be applied because they require sequence lengths of at least $10^6$ bits, and one other test (Maurer's) requires at least 387,840 bits and therefore can be applied only to sequence $\mathbf{x}_2$. Results are shown in Table~\ref{fig:nistresults}. The $p$-values for all applicable tests are found to exceed the chosen confidence threshold $\alpha=0.01$, and therefore the sequences can be considered random with a confidence of 99.9\%. 

To gather meaningful statistics in light of the chosen confidence threshold, one needs to subject at least $1/\alpha$ sequences to each test, since even for an ideal random source, a fraction $\alpha$ of sequences would be expected to fail the test, i.e., score $p < \alpha$. Therefore, next we let the NIST program break each sequence into $N=100$ nonoverlapping subsequences and run the tests on these subsequences. For three tests, our available sequence lengths impose constraints: (i) The Binary Matrix Rank and Maurer's tests cannot be applied to subsequences of $\mathbf{x}_1$ because of their insufficient length. (ii) For subsequences of $\mathbf{x}_2$, Maurer's tests is likewise inapplicable; moreover, the minimum sequence lengths required for an application of the  Binary Matrix Rank and Template Matching tests dictate that we may decompose $\mathbf{x}_2$ into only $N=20$ subsequences, implying a corresponding confidence threshold of only $\alpha=0.05$ (instead of $\alpha=0.01$). Test results are again given in Table~\ref{fig:nistresults}. The column labeled ``Proportion'' lists the number  of subsequences in the set that pass a test by scoring $p \ge \alpha$. The NIST suite considers the set of subsequences to pass the test if this proportion is larger than $n_\text{min} = 1-\alpha - 3\sqrt{ \alpha(1-\alpha)/N}$, where $N$ is the number of tested subsequences. For $\alpha=0.01$ and $N=100$, this gives $n_\text{min} =0.96$, i.e., at least 96 out of 100 sequences must pass the test. For $\alpha=0.05$ and $N=20$, we have $n_\text{min} =0.80$, i.e., the proportion must be at least 16/20. Our results show that the proportions of passing subsequences are all above these thresholds. The column labeled ``$P$-value'' in Table~\ref{fig:nistresults} represents the proximity of the distribution of $p$-values to a uniform distribution; passing a test corresponds to $P \ge 0.0001$ \cite{Rukhin:2010:ll}. All $P$-values are found to lie well above this threshold. In conclusion, our results show that the sequences $\mathbf{x}_1$ and $\mathbf{x}_2$ pass all applicable NIST tests.

\subsubsection{Borel normality}

Since statistical tests such as those of the NIST suite were originally developed to test PRNGs, they are not necessarily sensitive to the particular issues and properties that characterize QRNGs. QRNGs that pass the NIST tests may struggle with tests that probe aspects of incomputability and algorithmic randomness, or show at least no marked advantage over PRNGs in terms of their algorithmic properties \cite{Calude:2010:oo,Martinez:2018:kk,Abbott:2019:uu}. Here one particularly relevant and commonly used test is the Borel normality \cite{Calude:1994:hh,Calude:2010:kk}, which is a necessary (but not sufficient, e.g., Champernowne's constant \cite{Champernowne:1933:uu} is normal but computable) condition for algorithmic randomness and thus incomputability. It has been found to identify statistically significant differences between QRNGs and PRNGs, and the outputs of several experimental realizations of QRNGs have been shown not to be Borel normal \cite{Calude:2010:oo,Martinez:2018:kk,Abbott:2019:uu}. The likely cause of the failure of Borel normality has been attributed to experimental bias in the production of the bits \cite{Martinez:2018:kk,Abbott:2019:uu}. 

\begin{table}
\centering
\begin{tabular}{|c||c|cccc|}
  \hline
Sequence & Bound & $m=1$ & $m=2$ & $m=3$ & $m=4$ \\ \hline 
$\mathbf{x}_1$ & 0.0094 & 0.0011 & 0.0014 & 0.0020 & 0.0024 \\
$\mathbf{x}_2$ & 0.00495 & 0.00007 & 0.00090 & 0.00113 & 0.00109\\
\hline 
\end{tabular}
\caption{\label{fig:borel}Results of the Borel-normality test. The bound is given by $\sqrt{\log_2 \abs{\mathbf{x}}/\abs{\mathbf{x}}}$, where $\abs{\mathbf{x}}$ is the sequence length, with $\abs{\mathbf{x}_1}=2 \times 10^5$ and $\abs{\mathbf{x}_2}=8 \times 10^5$.}
\end{table}

Borel normality applied to a finite string measures whether all substrings of given length $m$ occur with the expected probability of $2^{-m}$ \cite{Calude:1994:hh,Calude:2010:kk}. It relates to the compressibility of the string by a lossless finite-state machine \cite{Ziv:1978:aa}. A string $\mathbf{x}$ of length $\abs{\mathbf{x}}$ is considered Borel normal if the following condition holds for all integer $m$ with $1 \le m \le \log_2 \log_2 \abs{\mathbf{x}}$ \cite{Calude:1994:hh,Calude:2010:kk}:
\begin{equation}\label{eq:borel}
\max_{1 \le j \le 2^m} \left| \frac{N_j^m(\mathbf{x})}{\abs{\mathbf{x}}/m} - \frac{1}{2^m} \right| \le \sqrt{\frac{\log_2 \abs{\mathbf{x}}}{\abs{\mathbf{x}}}},
\end{equation}
where $N_j^m(\mathbf{x})$ is the number of occurrences of the $j$th string drawn from the alphabet of all binary strings of length $m$. For our sequence lengths, the maximum value of $m$ is $m=4$. The results of the Borel-normality test applied to the two sequences $\mathbf{x}_1$ and $\mathbf{x}_2$ are shown in Table~\ref{fig:borel}. The given values correspond to the left-hand side of Eq.~\eqref{eq:borel} evaluated for $\mathbf{x}_1$ and $\mathbf{x}_2$. Both sequences comfortably pass the test, as all values are well below the bound given by the right-hand side of Eq.~\eqref{eq:borel}. Note that for $m=1$, the Borel-normality condition amounts to assessing the relative frequencies $p_0$ and $p_1$ of 0s and 1s. For our two sequences, this bias is extremely small: For $\mathbf{x}_1$ we have $\abs{p_0-0.5}=1.1 \times 10^{-3}$, and for $\mathbf{x}_2$ we have $\abs{p_0-0.5}=7.0 \times 10^{-5}$. 

As an additional check, we also estimate the algorithmic complexity of the sequences $\mathbf{x}_1$ and $\mathbf{x}_2$ by calculating their compressibility in terms of information density \cite{Hamming:1980:aa} using the program \textsc{ent} \cite{ent}. We find information densities per bit of 0.998767 and 0.999739 for $\mathbf{x}_1$ and $\mathbf{x}_2$, respectively. These values are very close to the maximum of 1. This confirms that the sequences are virtually incompressible, indicating a large amount of randomness.

\section{\label{sec:discussion}Discussion}

Our approach of using the parity of photon counts produces bias-free random numbers that pass the applicable statistical tests in the NIST suite and are Borel normal, without the need for an unbiasing postprocessing procedure or time-tagging. As we obtain the photon counts from a measurement of the CHSH violation for a polarization-entangled state, we can simultaneously confirm the nonclassical nature of the photon statistics and use the measured violation to estimate the min-entropy of the source. While a rigorous, loophole-free \cite{Christensen:2013:aa,Shalm:2015:oo} demonstration of such certified randomness \cite{Pironio:2010:aa} is beyond the aim and scope of our experiment, the use of the measured CHSH violations may be regarded as a concurrent benchmark for assessing the randomness of the process and for complementing statistical and algorithmic tests of the output. The min-entropy bound we obtain from quantum state tomography serves as an additional confirmation of the presence of quantum randomness in the measured photon events. In this way, we may infer that the photon count statistics we measure are not just due to classical fluctuations, but are, at least in part, due to a genuinely quantum-mechanically random process without classical analog. 

An obvious disadvantage of our method of generating bits from the parity of photon counts is its relative inefficiency, since we obtain just one number from a collection of photon events (namely, the number of counted photons in a given time interval $\tau$) and extract only the least significant bit from that number. One effective way of substantially increasing the bit rate would be to make the counting interval $\tau$ shorter. While the particular coincidence counting unit we use \cite{Lord:noyear:uu} does not allow for counting intervals shorter than $\tau=\unit[0.1]{s}$ (limiting the bit rate to around \unit[20]{Hz}), even relatively simple other units \cite{Branning:2009:aa,Branning:2010:oo} provide the choice of much smaller intervals, and therefore the possibility of significantly higher bit rates. Another approach to boosting bit rates would be to use additional bits of the photon counts.  

We note here that if time-tagging capabilities are available, one could alternatively apply the parity method for bit generation to the photon arrival times, rather than the photon counts, such that each photon event generates one bit. This was done, for example, in Ref.~\cite{Martinez:2018:kk}, where the resulting sequence was shown to be Borel normal. In such an experiment, however, a complementary verification of randomness via a min-entropy estimation from a CHSH violation or state tomography, as we have performed it, would not be possible in the same way, since the bits are not generated from measurements of polarization, which is the observable relevant to these types of min-entropy estimation. 

Because the polarization-entangled state we generate implies the presence of correlations, the coincidence events $AB$, $A'B$, $AB'$, and $A'B'$ are not independent and one might be concerned that such correlations could diminish the randomness of the sequence $\mathbf{x}_2$ produced from all four of these coincidences. This would indeed be the case if we were to generate each bit from individual measurement events rather than from parity, and a sequence experimentally produced in this way has been shown to fail the NIST tests \cite{Pironio:2010:aa}. The sequences $\mathbf{x}_1$ and $\mathbf{x}_2$, however, both pass all of our randomness tests. Thus, within the scope of the tests we have performed, the Bell correlations do not have a detectable influence on the randomness of the sequence $\mathbf{x}_2$, indicating that the effect of such correlations becomes effectively washed out when considering the parity variable of accumulated counts.

\begin{acknowledgments} 
This work was supported by the M.~J. Murdock Charitable Trust  (NS-2015298) and by the SURE program of the University of Portland.
\end{acknowledgments}


%

\end{document}